\documentclass[amsmath,amssymb,longbibliography,
aps,prl,reprint]{revtex4-2}
\usepackage{graphicx,color}
\usepackage[normalem]{ulem}
\usepackage{xcolor}
\usepackage{dcolumn}
\usepackage{comment}
\usepackage{bm}
\usepackage{cases}  
\usepackage{upgreek}
\definecolor{colortodo}{RGB}{255,0,0}
\definecolor{colortodo2}{RGB}{0,0,255}

\usepackage{tabularx}

\def\beq{\begin{equation}}
\def\eeq{\end{equation}}
\def\bea{\begin{eqnarray}}
\def\eea{\end{eqnarray}}

\usepackage{natbib}
\usepackage{hyperref}

\begin{document}



\title{Elasto-Hydrodynamic Propulsion of a Magnetically Actuated Filament}

\author{Sohum Kapadia, Julien Chopin, Arshad Kudrolli}
\affiliation{$^{1}$Department of Physics, Clark University, Worcester, Massachusetts 01610, USA\\
$^{2}$Instituto de F\'isica,  Universidade Federal da Bahia, Salvador-BA 40170-115, Brazil}
\date{\today}

\begin{abstract}
We investigate the low-Reynolds-number propulsion of a slender elastic filament with a dipolar magnetic head actuated by an oscillating field in a viscous fluid by studying its strokes and net forward motion. To capture these dynamics, we employ an elasto-hydrodynamic (EH) framework that couples Euler-Bernoulli beam mechanics with resistive force theory. Unlike prescribed-kinematics models, filament shapes here emerge self-consistently from the actuation and the force and torque boundary conditions (BCs).  We demonstrate that viscous boundary contributions are crucial for quantitative agreement and show that the swimming dynamics are governed by the EH length and a magneto-viscous-elastic stroke amplitude introduced here. The swimming speed is non-monotonic with increasing ratio of the swimmer length to the EH length, and is shown to reach a maximum when the swimmer length is on the order of the EH length.  We further discuss the analytical limit in which the tail BCs can be described as free, and the limitations that arise when viscous contributions to the BCs are ignored. 
\end{abstract}

\maketitle

Locomotion of soft swimmers at low Reynolds number is governed by a subtle interplay between elasticity and anisotropic viscous drag, with classical foundations laid by Gray and Hancock and later by Lighthill \cite{gray1955movement,Lighthill1975,wiggins1998trapping,wiggins1998flexive}. Resistive force theory (RFT) has provided an effective description of propulsion by slender filaments in this regime, while slender-body theory (SBT) refines the hydrodynamics by accounting for nonlocal interactions along the filament \cite{gray1955propulsion,Lighthill1975,Johnson1980}. Modern syntheses emphasize the conditions under which these approaches converge, particularly when integrated into elasto-hydrodynamic (EH) models that incorporate Euler--Bernoulli beam theory to predict both emergent swimmer shapes and swimming speeds~\cite{wiggins1998flexive,lauga2009hydrodynamics,lauga2020fluid,pak2011high,yu2006experimental}. Yet, even within linear elasto-hydrodynamics, quantitative discrepancies between predicted and measured waveforms and speeds have persisted~\cite{rodenborn2013propulsion,liu2025effective,htet2025load}. These differences are frequently attributed to the inherent approximations of RFT, but they may also originate from the physical closure of the problem at the boundaries rather than a failure of the bulk hydrodynamic description itself. 

Experiments with a pinned filament with prescribed amplitudes at low Reynolds number demonstrated that linear EH models can capture its shapes and propulsive forces, despite the use of idealized boundary conditions that neglect viscous end-point effects~\cite{yu2006experimental}. Similarly, RFT has been shown to provide accurate speed predictions by utilizing a periodic swimming sheet geometry where end-point singularities are intrinsically avoided~\cite{Dasgupta2013}. Agreement has also been obtained by fitting drag anisotropy to prescribed waveforms, treating the observed kinematics as a model input rather than a dynamical output~\cite{Johnson1979,Friedrich2010,Kantsler2012}. Furthermore, numerical simulations of magnetically actuated rigid, helically shaped microswimmers with attached payloads have found that their location and aspect ratios can significantly affect the optimal shape for swimming which cannot be predicted using RFT by itself~\cite{Keaveny2013}. Consequently, resistive-force-based descriptions of freely swimming elastic bodies where the waveforms are not prescribed but emerge self-consistently alongside finite boundary effects remain an open question.

Here we investigate a magnetically actuated, freely swimming elastic filament in a viscous fluid, combining precision experiments with a linear EH theory coupling Euler--Bernoulli beam mechanics to RFT. In contrast to prescribed-kinematics approaches, filament shapes in our system emerge self-consistently from the specified actuation strength, with force and torque boundary conditions inferred directly from measured shapes. Refined boundary conditions are constructed to account for the viscous drag and the physical geometry of the head and tail besides the magnetic torque. We show that discrepancies between theory and experiment for finite swimmers are resolved by addressing incomplete boundary closure, thereby demonstrating that bulk RFT remains quantitatively valid when viscous force and torque contributions at the boundaries are properly incorporated. We show that the swimming dynamics are governed by the elasto-hydrodynamic length and the magneto-viscous-elastic stroke amplitude.  Finally, we analyze the limits where hydrodynamic contributions to the BCs can be neglected.


\begin{figure}
\centering
\includegraphics[width=0.9\linewidth]{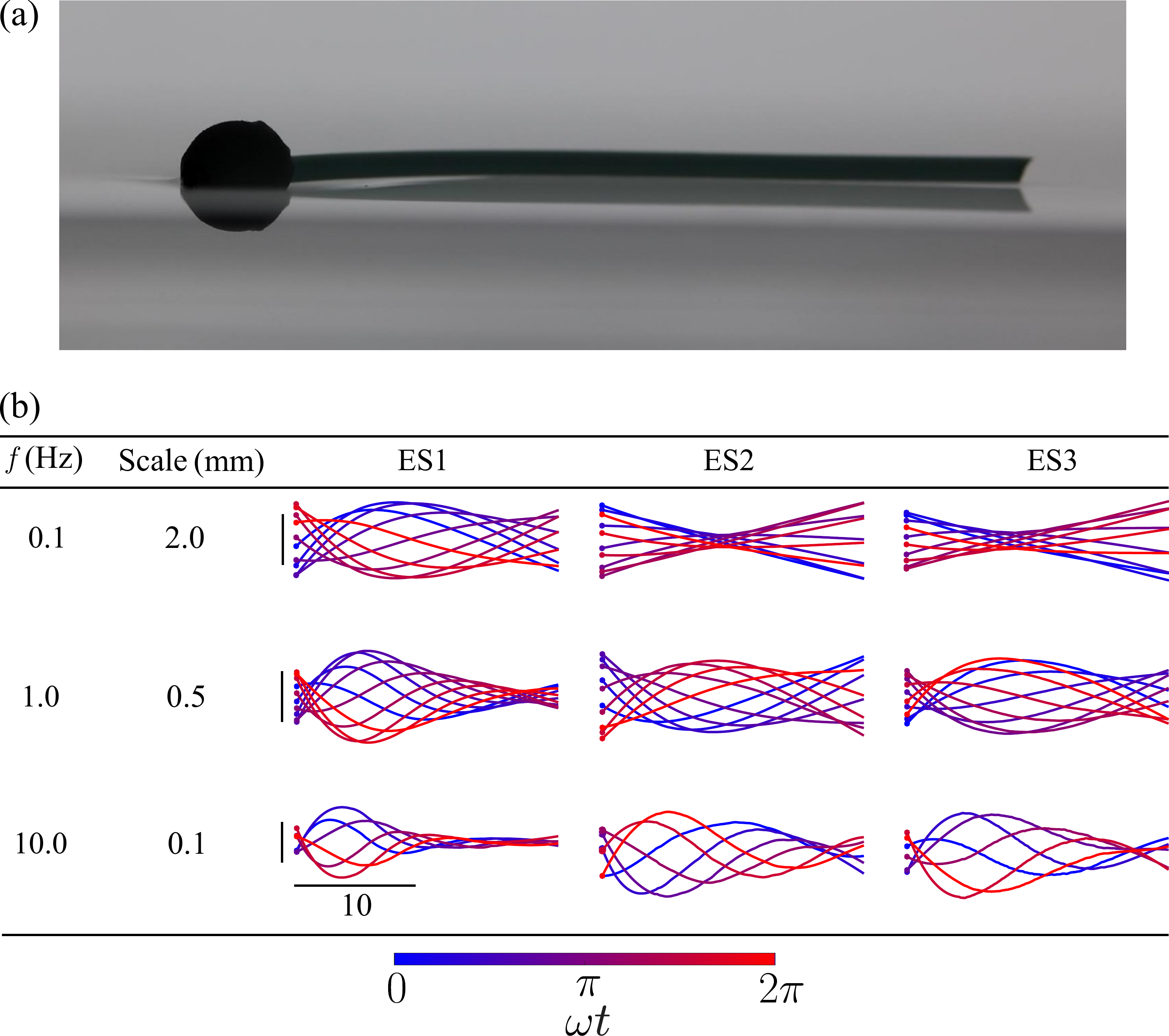}
\caption{(a) An image of the magneto-elastic swimmer showing the swimmer at a glycerin-silicone oil interface. (b) The tracked body shapes plotted over one oscillation time period $\omega t$ for three swimmers at various driving frequencies in the co-moving reference frame. The head is denoted with a solid black marker. The snapshots are at time interval $\Delta t = 16.7$\,ms.}
\label{fig:experimental setup}
\end{figure}

Figure~\ref{fig:experimental setup}(a) shows a magneto-elastic swimmer composed of a 3D printed spherical head of diameter $d_h = 3$\,mm, with an embedded neodymium magnet with magnetic moment $m = 1.14 \times10^{-3}$\,Am$^2$, attached co-linearly with a cylindrical polyvinyl-siloxane (PVS) filament of length $L=21.5$ mm and diameter $d = 0.8$ mm. The density of the filament $\rho_t = 0.97 \textrm{ g\,cm}^{-3}$, and the head is designed with an air cavity such that the composite  density of the swimmer is $\rho_s = 1.05 \pm 0.05 \textrm{ g\,cm}^{-3}$. This system builds on and complements previous work with magnetoelastic filaments at high Reynolds numbers~\cite{ramananarivo2013passive,biswas2023dynamics}. Swimmers ES1, ES2, and ES3 were fabricated from Zhermack Elite Double silicones with increasing Shore hardness values of 8, 22, and 32, respectively, to systematically vary elasticity. An acrylic transparent container ($20 \times 10 \times 8$ cm$^3$) filled with glycerin (dynamic viscosity $\eta_g = 1.1 \pm 0.1 \textrm{ Pa\,s }$, density $\rho_g = 1.26\textrm{ g\,cm}^{-3}$) and silicone oil ($\eta_o = 0.1$\,Pa\,s and $\rho_o =  0.97\textrm{ g\,cm}^{-3}$). Fig.~\ref{fig:experimental setup}(a) shows an image of the swimmer floating at the interface formed with silicone oil on top and glycerin at the bottom. We obtain an effective viscosity $\eta \simeq 0.4$\,Pa\,s, at the interface, which is higher than the bulk viscosity of the Si-oil with complementary measurements (see \cite{SI}). A spatially uniform oscillating magnetic field $B(t) = B_0 \cos(\omega t)$, where $B_0$ is the field strength, $\omega = 2\pi f$ with $f$ being the applied frequency, and time $t$, is generated using Helmholtz coils connected to an amplifier and a function generator. 
The resulting magnetic torque $M(t) = M_0\cos(\omega t)$ drives head oscillations which propagate along the elastic tail of the swimmer. Assuming small-slope deflections, $M_0 \simeq m B_0$. 


The swimmer is imaged from above against a bright illumination to enhance the contrast and enable full body tracking of the robot using a Canon EOS R5 camera. Figure~\ref{fig:experimental setup}(b) shows the tracked body shapes over a single oscillation period $T$ for the 3 elastic swimmers at $f =$ 0.1 Hz, 1.0 Hz, and 10 Hz in a co‑moving reference frame. The swimmer exhibits only weak bending at the lowest frequency for the stiffest filament, but undergoes progressively larger deformations with increasing driving frequency and decreasing bending rigidity. 

\begin{figure}
    \centering
    \includegraphics[width=\linewidth]{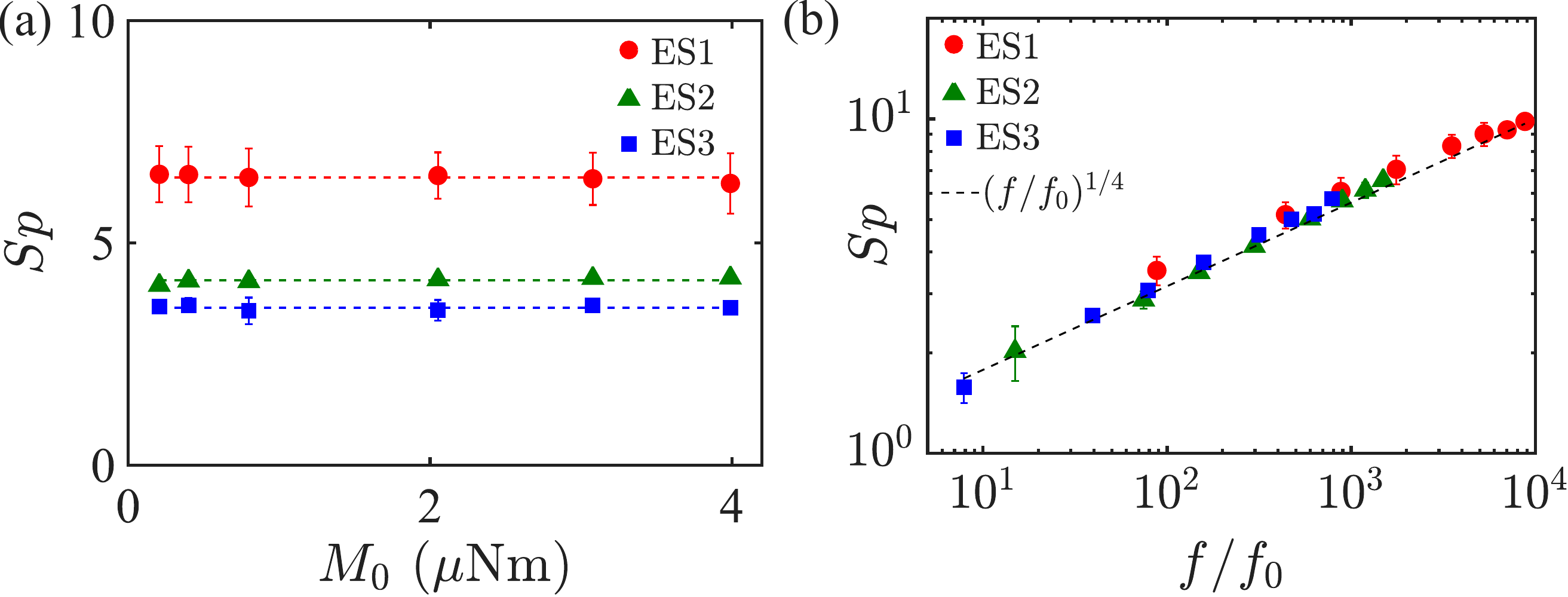}
    \caption{(a) Comparison of measured $Sp$ as a function of $M_0$ ($f = 2$\,Hz) (b) $Sp$ versus $f/f_0$ ($M_0 \simeq 0.4$ $\mu$Nm) is well described with no further adjustable parameters. $f_0$ is determined for each elasticity from the constant fit in (a), using Eq.~\ref{eq:f_0}.  }
    \label{fig:Sp_estimation}
\end{figure}

Assuming the dynamics correspond to the small-amplitude, low-Reynolds-number regime, the filament shape is governed by the EH equation~\cite{wiggins1998trapping},
\begin{equation}
-K \frac{\partial^4 h}{\partial x^4} - \zeta \frac{\partial h}{\partial t} = 0,
\label{eq:ehd}
\end{equation}
where $K = E I$ is the bending stiffness with the Young's modulus $E$ and area moment of inertia  $I$, and the perpendicular drag coefficient of the filament $\zeta = 4\pi\eta/[\ln(L/d) + c]$, as given in \cite{wiggins1998flexive} with $c = 0.5$ \cite{Yang2017}.

We define dimensionless variables $X = x/\ell_{ev}$, $T = \omega t$, and $H = h/h_a$, where $\ell_{ev} = (K/\omega \zeta )^{1/4}$ is the elasto-hydrodynamic length \cite{wiggins1998trapping}, and the corresponding dimensionless Sperm number,  
\begin{align}
\label{eq:Sp}
    Sp = \frac{L}{\ell_{ev}} = L\left(\frac{\omega\zeta}{K}\right)^{1/4}.
\end{align} 
Further, we introduce the magneto-viscous-elastic amplitude $h_a = M_0 / \sqrt{\zeta \omega K}$ {which arises from balancing the driving torque $M_0$ against the resisting elasto-hydrodynamic torque that scales as $K \kappa$, where $\kappa = \partial^2h/\partial x^2 \sim h_a/\ell_{ev}^2$ is the typical curvature of the filament} (also see SI).

With a complex harmonic solution $H(X,T) = \psi(X) e^{i T}$, the equation reduces to an ordinary differential equation for the spatial envelope $\psi(X)$:
\begin{equation}
\frac{d^4 \psi}{d X^4} + i\psi = 0,
\end{equation}
with the general solution expressed as:
\begin{equation}
\label{eq:EHD_sol}
\psi(X) = c_1 \cosh(kX) + c_2 \sinh(kX) + c_3 \cos(kX) + c_4 \sin(kX),
\end{equation}
where $k = e^{-i\pi/8}$ is the complex wave number, $c_j$ with $j=1,2,3,4$ are coefficients which depend on boundary conditions corresponding to head located at $X =0$ and tail end located at $X = Sp$ after rescaling lengths~\cite{wiggins1998trapping}. 

We determine $Sp$ and the coefficients $c_j$ by fitting Eq.(\ref{eq:EHD_sol}) to the experimentally tracked shapes of the undulating elastic filaments. (The fitted forms are shown in the \cite{SI} and are in excellent agreement.) We plot $Sp$ as a function of $M_0$ for the three swimmers in Fig.\,\ref{fig:Sp_estimation}(a), and observe that it remains constant while being systematically higher for softer filaments. It is possible to calculate $Sp$ directly from Eq.(\ref{eq:Sp}) if the physical constants are known. Because PVS swells somewhat when immersed in Si-oil, changing $K$, we obtain $K$ for each swimmer {\it in situ} by using the measured $Sp$ and known $L,\omega,\zeta$ instead of Eq.~(\ref{eq:Sp}). We find $K = 1.68$\,nNm$^2$, 12.00\,nNm$^2$, and $19.30$\,nNm$^2$ for ES1, ES2, and ES3, respectively. Then, these calibrated $K$ are used in calculating $Sp$ as a function of frequency using Eq.~(\ref{eq:Sp}), and comparing with the directly measured $Sp$ versus $f/f_0$ log-log plot shown in Fig.\,\ref{fig:Sp_estimation}(b), where $f_0$ is an EH frequency scale defined as,
\begin{equation}
\label{eq:f_0}
f_0 = \frac{K}{2 \pi \zeta L^4}.
\end{equation}
The data collapse onto a line with 1/4 slope according to Eq.~(\ref{eq:Sp}) without any further adjustable parameters, thereby validating the linear EH framework used to describe our swimmers.


\begin{figure}[h!]
    \centering
     \includegraphics[width=0.99\linewidth]{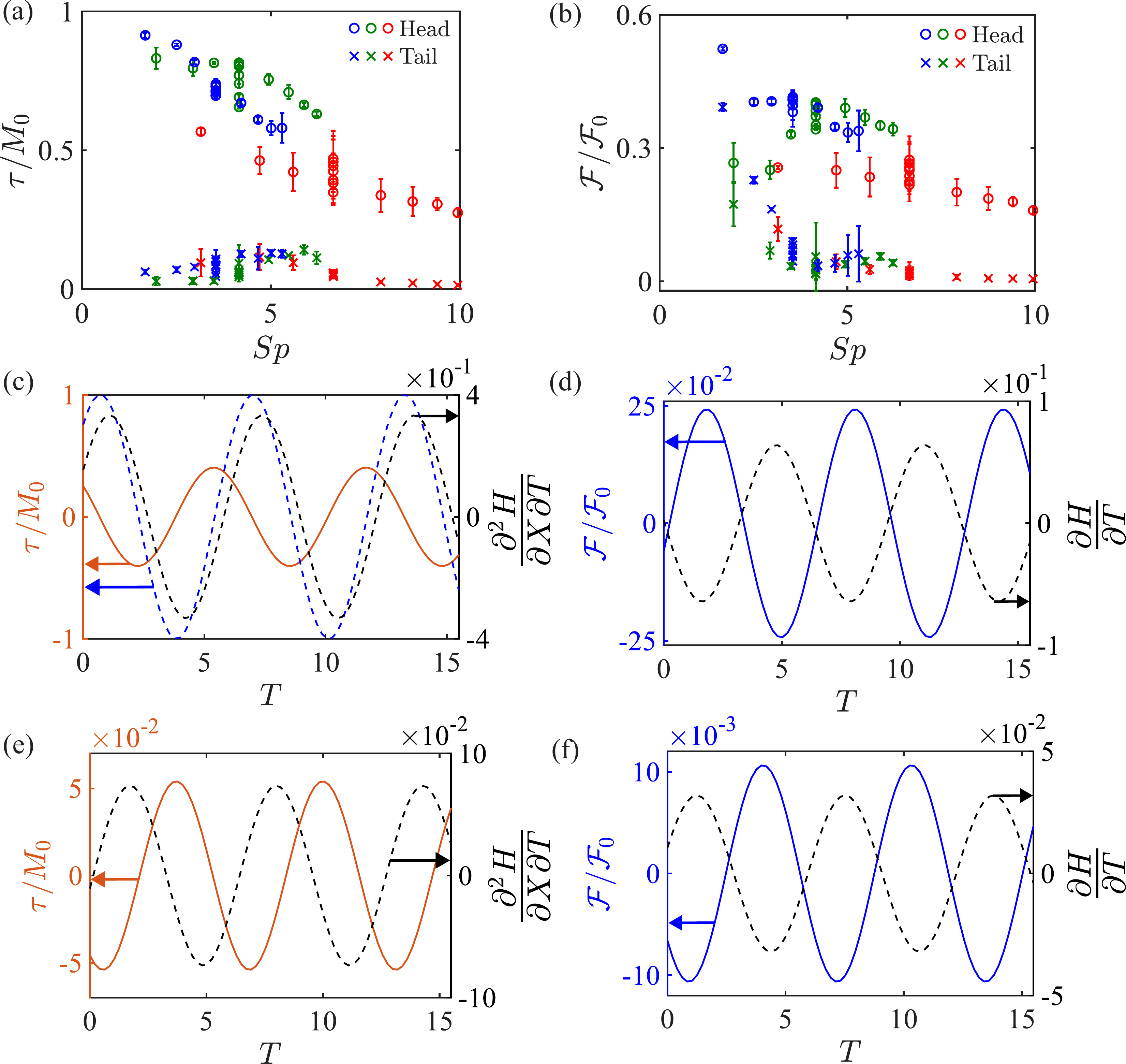}
     \caption{(a,b) The boundary torque {\Large $\tau$} and force $\mathcal{F}$ normalized by the driving torque $M_0$ and the characteristic EH force $\mathcal{F}_0 = M_0/ \ell_{ev}$. (c-f) {\Large $\tau$}$/M_0$ and $\mathcal{F}/\mathcal{F}_0$ plotted as a function of $T$ corresponding to the magnetic and elastic torque at the head and tail. The velocity of the head, tail, and their angular speeds are also plotted.} 
    \label{fig:abs_BC}
\end{figure}

To determine the appropriate BCs, the torque {\Large $\tau$} $= M_0 \frac{\partial ^2 H}{\partial X^2}$ and force $\mathcal{F} = \mathcal{F}_0 \frac{\partial^3H}{\partial X^3}$ experienced are obtained by evaluating the second and third derivatives of $H(X,T)$, using the experimentally obtained $c_j$ at $X=0$ (head) and $X = Sp$ (tail), where the characteristic scales for the torque and force are $M_0$ and $\mathcal{F}_{0} = M_0/\ell_{ev}$, respectively. We plot the normalized torque {\Large $\tau$}$/M_0$ and force $\mathcal{F}/\mathcal{F}_0$ as functions of $Sp$ in Fig.~\ref{fig:abs_BC}(a) and Fig.~\ref{fig:abs_BC}(b), respectively for ES1 at $f =2$~Hz, and $M_0 = 0.4$~$\mu$Nm, as reference. We find that they are significantly higher at the head than at the tail because of its larger size. While decreasing with $Sp$, they cannot be {\it a priori} assumed to be negligible at the tail.

Fig.~\ref{fig:abs_BC}(c)  shows plots of {\Large $\tau$}$/M_0$  (solid red line), the normalized applied torque $M(t)/M_0$ (dashed blue line), and the rate of change of the slope $\frac{\partial}{\partial T}\left(\frac{\partial H}{\partial X}\right)$ (dashed black line) as functions of normalized time $T$, at $X=0$ to infer the various contributions to BCs. As can be seen from Fig.~\ref{fig:abs_BC}(c), the net torque experienced by the filament is lower than the applied torque and decreases even further with higher $Sp$. To account for this, we add a viscous drag term proportional to the rate of change of the slope. Thus,
\begin{equation}
\frac{\partial^2 H}{\partial X^2} = e^{i(T + \phi_0)} - \alpha_H\xi_{H} \frac{\partial}{\partial T}\left(\frac{\partial H}{\partial X}\right),
\label{eq:BC1}
\end{equation}
where $\phi_0$ is the relative phase of the applied torque required to satisfy Eq.~(\ref{eq:BC1}). Here, $\xi_H = \zeta_H d_h/(\zeta\,\ell_{ev})$, where $\zeta_H$ is the drag coefficient of the spherical head, and $\alpha_H$ is a constant of order unity. Further, we plot $\mathcal{F}/\mathcal{F}_0$ and $\frac{\partial H}{\partial T}$ with time in Fig.~\ref{fig:abs_BC}(d), and observe that the force is essentially $\pi$ out-of-phase with the head velocity. Hence, we formulate the force boundary condition at $X= 0$ as the balance between elastic bending and viscous drag,  
\begin{equation}
   \frac{\partial^3 H}{\partial X^3} = - \xi_H \frac{\partial H}  {\partial T} \,. 
    \label{eq:BC2}
\end{equation}

Noting that the phase relation between the elastic term and the kinematic term is not strictly $\pi$ out-of-phase in Fig.~\ref{fig:abs_BC}(e,f), we need to consider additional reactance terms in determining BCs at the tail. Accordingly, we assume that the torque BC at the tail,
\begin{equation}
\frac{\partial^2 H}{\partial X^2} =  - \alpha_T \xi_T \frac{\partial}{\partial T} \left( \frac{\partial H}{\partial X} \right)- \Gamma_{RT} e^{i\phi_{RT}},
\label{eq:BC3}
\end{equation}
and the force BC at the tail,
\begin{equation}
   \frac{\partial^3 H}{\partial X^3} = -\xi_T \frac{\partial H}  {\partial T} - \Gamma_T e^{i\phi_T}, 
    \label{eq:BC4}
\end{equation}
where $\xi_T = \zeta_{T}d/\zeta \ell_{ev}$ is the non-dimensional viscous drag coefficient at the tail of diameter $d$, and $\Gamma_T$, $\Gamma_{RT}$, $\phi_T$ and $\phi_{RT}$ correspond to reactance terms that account for the additional effects that are not accounted by the viscous drag terms alone due to the fact that the swimmer moves at the oil-glycerin interface. At each $f$, we obtain the drag coefficients $\xi_{H}, \xi_T, \alpha_{H}$ and $\alpha_T$ across all the experiments. (Their plots as a function of experimental parameters $M_0$ and $f$ for all three swimmers can be found in SI.) As we infer the BCs in Eq.(\ref{eq:BC1}-\ref{eq:BC4}) from the experimentally obtained $c_j$, we identify these as the observed BCs.  

\begin{figure*}
    \centering
    \includegraphics[width=0.85\linewidth]
    {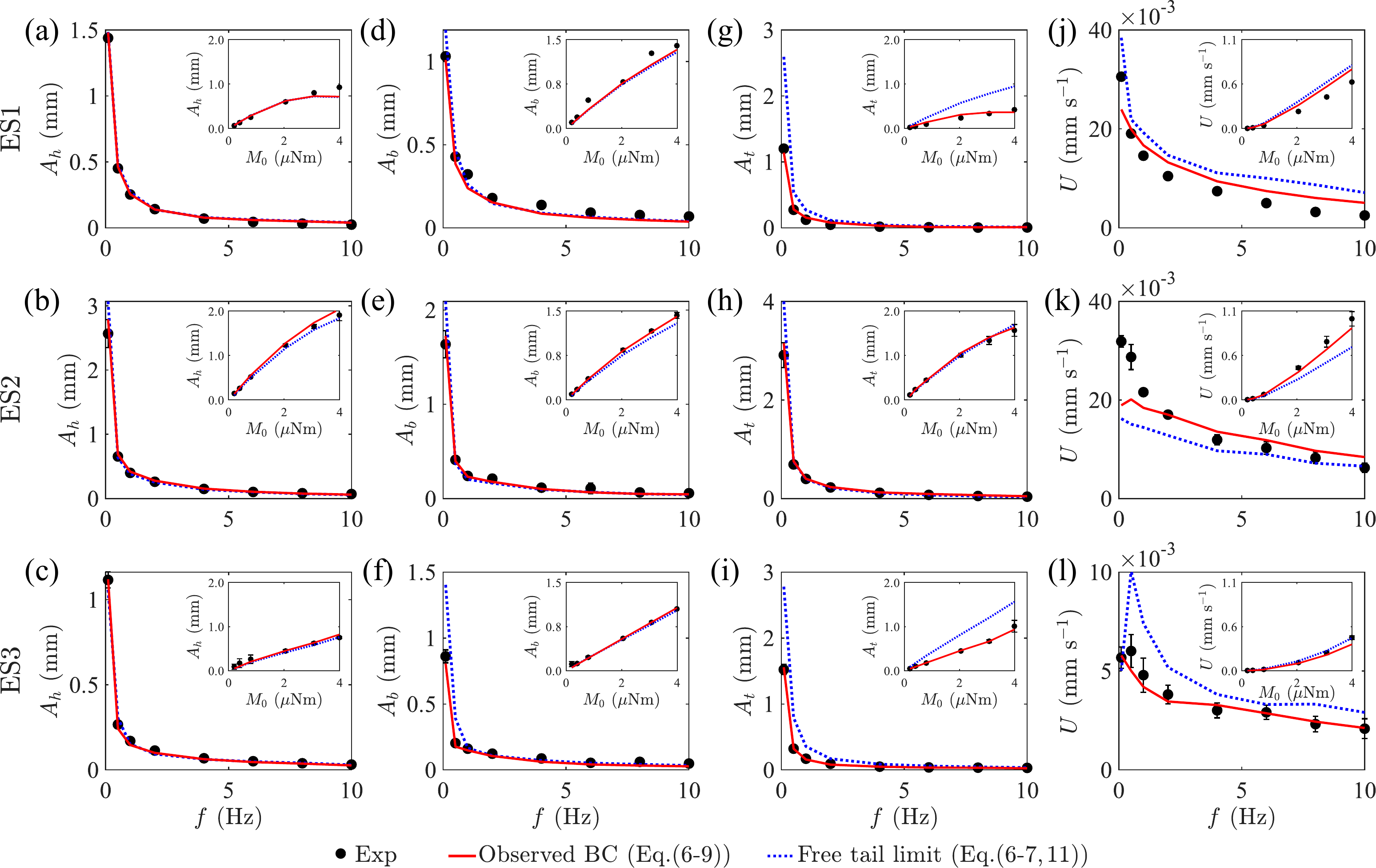}
    \caption{Comparison of the amplitudes (a-i) and speeds (j-l)  for all three swimmers, between the experiment, the EH model using observed BC Eq.(\ref{eq:BC1}-\ref{eq:BC4}) (solid red line) and in the free tail limit (dashed blue line). Insets: Comparisons as a function of $M_0$.} 
    \label{fig:Exp_vs_RFT_all_models}
\end{figure*}

We plot the head amplitude $A_h$, the tail amplitude $A_t$ and the body amplitude $A_b$ to compare the body strokes observed in the experiments (solid black bullets) and the EH model (solid red line) in Fig.~\ref{fig:Exp_vs_RFT_all_models} with $f$ and $M_0$ (in the Insets). The measured amplitudes decreases with increasing driving frequency $f$ and increases with the driving torque $M_0$ as captured by the amplitude scale $h_a = M_0/\sqrt{\zeta \omega K}$. They are in excellent agreement with minor deviations.

We now use RFT to determine their mean propulsion speed, which assumes that the net force experienced by the filament can be considered as a superposition of infinitesimal forces on the elements along the filament. The mean forward speed $\langle U \rangle$ is obtained by the balance between the time integrated thrust generated along the filament with time integrated total drag experienced by the swimmer (see \cite{SI}). The total drag on the swimmer is the combination of the hydrodynamic drag on the filament, spherical head and an additional drag due to the viscous interface. As clearly seen in Movie S2, the interface is deformed around the head and tail, leading to end-point drag forces.
\begin{equation}
\begin{gathered}
\langle U \rangle = \omega h_a(1-\beta) \\ 
   \frac{ \displaystyle \int_0^{Sp} \int_0^{2\pi} \dot{H}\sin\theta\cos\theta \, dT \, dX}
    {\displaystyle \int_0^{Sp} \int_0^{2\pi} \big[\sin^2\theta + \beta\cos^2\theta\big] dT\,dX + \displaystyle \int_0^{2\pi} \xi_B \ dT},
    \end{gathered}
    \label{eq:U0}
\end{equation}
where $\dot{H} = \partial H/ \partial T$, $\theta = \tan^{-1}(h_a\partial H/\ell_{ev}\partial X)$, $\beta = \zeta^\parallel/ \zeta$ is the drag anisotropy of the elastic tail, here assumed to be $\beta  = 0.5$ for a slender filament, and  $\xi_B = (\zeta_B d_h/\zeta \ell_{ev})$ is the effective drag coefficient due to the spherical head and the interfacial boundary effects. We determine the $\zeta_B$ by calibrating the mean propelling speeds obtained from the Eq.~(\ref{eq:U0}), using the fitted shapes to the measured mean propelling speeds. Using elastic swimmer ES2, and the speed with varying $M_0$ as the reference data, we obtain $\zeta_B/3\pi d_h \simeq 1$\,Pa\,s. Because of the cumulative effect of the spherical head and the interface, the swimmer thus experiences higher drag compared to a sphere of diameter $d_h$.

We calculate the mean propelling speed $\langle U \rangle$ for all three swimmers using the RFT model using Eq.(\ref{eq:U0}), where $H(X, T)$ corresponds to the shapes obtained from EH model with the observed BCs. We compare them with the experimentally measured speeds in Fig.~\ref{fig:Exp_vs_RFT_all_models}(i-l) and their insets while varying $f$ and $M_0$, respectively. The measured speed overall decreases with increasing frequency and increases non-monotonically with $M_0$. The effect of bending stiffness $K$ on the speed is non-linear, as ES2 swims the fastest followed by ES1 and the ES3 being the stiffest and slowest. The speeds obtained with RFT model and experiments are in excellent agreement with deviations observed at lowest frequency, where the amplitudes are not strictly small. 

We examine the limit where the tail BCs can be considered as torque and force free, i.e. at the tail $X=Sp$,
\begin{gather}
   \frac{\partial^2 H}{\partial X^2}  = 0,\, 
   \, {\rm and}\,\,
   \frac{\partial^3 H}{\partial X^3} = 0 \label{eq:BC4_free}.
\end{gather}

Keeping Eq.(\ref{eq:BC1}-\ref{eq:BC2}) at the head ($X=0$), the coefficients $c_j$ can be analytically obtained by solving the following set of four linear equations:
\begin{align}
\begin{split}
c_1 &= \big[ e^{i\phi_0} ( \sin(k)\sinh(k) + \cos(k)\cosh(k)\\
    &\quad + D\cos(k)\sinh(k)  - D\cosh(k)\sin(k) \\
    &\quad - 1 ) \big] \Big/ \Delta,
\\
c_2 &= - \big[ e^{i\phi_0} ( D + \cos(k)\sinh(k) + \cosh(k)\sin(k) \\
    &\quad + D\cos(k)\cosh(k) - D\sin(k)\sinh(k) ) \big] \Big/ \Delta,
\\
c_3 &= \big[ e^{i\phi_0} ( \sin(k)\sinh(k) - \cos(k)\cosh(k) \\ 
    &\quad- D\cos(k)\sinh(k) 
     + D\cosh(k)\sin(k)\\ 
     &\quad+ 1 ) \big] \Big/ \Delta,
\\
c_4 &= - \big[ e^{i\phi_0} ( D + \cos(k)\sinh(k) + \cosh(k)\sin(k) \\
    &\quad + D\cos(k)\cosh(k) + D\sin(k)\sinh(k) ) \big] \Big/ \Delta,
\end{split}
\end{align}
where,
\begin{align*}
\begin{split}
\Delta &= 2 ( CD - k^2 + C\cos(k)\sinh(k) \\
    &\quad + C\cosh(k)\sin(k) + k^2\cos(k)\cosh(k)\\
    &\quad - Dk^2\cosh(k)\sin(k) + CD\cos(k)\cosh(k)\\
    &\quad+ Dk^2\cos(k)\sinh(k)),\\ 
D &= \xi_H k^4, \,\, {\rm and} \,\, 
C = \alpha_H\xi_Hk^5.
\end{split}
\end{align*}
We substitute the calculated coefficients in the Eq.(\ref{eq:EHD_sol}) to get the spatial envelope $\psi(X)$. By multiplying it with the temporal part $e^{iT}$, we recover the full solution to Eq.(\ref{eq:ehd}). The real part of the obtained analytical solution represents the form of the swimmer with imposed BC at head and free tail.

To compare the shapes obtained from the free tail limit with the tracked shapes, we plot them adjacent to each other across the frequency range ($f = 0.1$Hz, $1$Hz and $10$Hz) and for three swimmers (ES1, ES2, ES3) with varying $K$ (see \cite{SI}).
The shapes from the free tail limit captures the trends with varying $f$ and $K$, evolving from rod-like oscillations at lowest frequency and stiffest swimmer to increasingly floppy oscillations at higher frequency and softer swimmer, closely following the tracked shapes. While the waveforms are in overall agreement, the free tail limit leads to shapes with higher tail deflections (see Fig.~\ref{fig:Exp_vs_RFT_all_models}(g-i)). This is expected because ignoring the forces and torques at the tail leads to higher deflection there. The speeds obtained from the shapes in the free tail limit Fig.\ref{fig:Exp_vs_RFT_all_models}(j-l), shows a striking agreement in the overall speeds, similarly deviating at lower frequencies. Thus, while the speeds obtained using RFT and the full observed BCs show good agreement, the ones calculated assuming free tail BCs, while deviating somewhat, also give a good description of the observed trends with $f$ and $M_0$. 

\begin{figure}
    \centering    
    \includegraphics[width=0.7\linewidth]{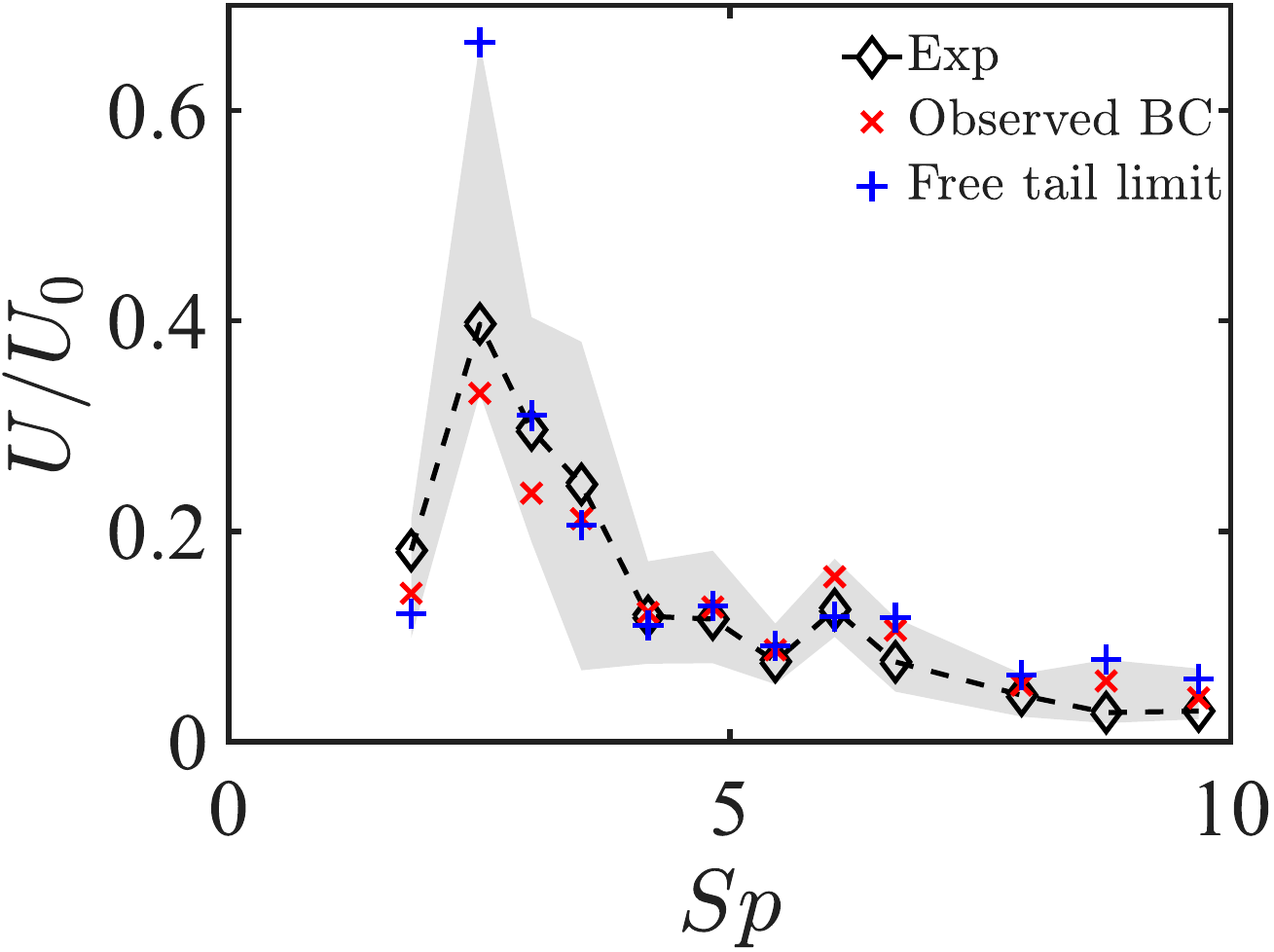}
    \caption{ The normalized propulsion speed $U/U_0$ as a function of $Sp$
    showing a non-monotonic dependence with a maximum propulsion at $Sp \approx 2$, indicating optimal propulsion when the filament length is comparable to the EH length.
    The observed $U/U_o$ is in agreement with those obtained with experimentally inferred boundary conditions and the free-tail limit.}
    \label{fig:UoSp}
\end{figure}

Finally, we obtain the normalized swimming speed $U/U_0$ as a function of $Sp$, where $U_0 = \omega A_b^2/\ell_{ev}$ corresponds to the speed scale using the applied frequency, the square of the  body amplitude,  and  the EH decay length. Figure~\ref{fig:UoSp} shows the speeds versus $Sp$ averaged across frequency and swimmers in unit-$Sp$ bins to illustrate the overall trends.  We observe that $U/U_0$ increases from the stiff limit $Sp \approx 0$, reaches a peak at $Sp \approx 2$ before decaying with $Sp$. This overall non-monotonic dependence and the existence of a peak are similar to those reported by Lauga~\cite{Lauga2007} for actuated elastic filaments with prescribed shapes. Our system also shows a consistent trend but further under self-consistent amplitudes and boundary conditions.  Comparing the calculated speeds using RFT using the experimentally inferred boundary conditions, we find agreement within experimental scatter. Good agreement is also observed in the free-tail limit except near peak-$Sp$.  

Thus, we have demonstrated that the propulsion of magnetically actuated elastic swimmers at low Reynolds number is governed by a coupled elasto-hydrodynamic equation in which the filament shape, amplitude, and swimming speed emerge self-consistently from the actuation and boundary conditions. By incorporating viscous force and torque contributions at both ends of the filament, we resolve longstanding ambiguities between experiment and resistive-force-based predictions for finite swimmers. Our results show that the swimming speed exhibits a clear optimum when the filament length is comparable to the elastohydrodynamic length, reflecting the competition between elastic deformation and viscous damping. The collapse of the data across a wide range of driving conditions demonstrates that resistive force theory, when supplemented with physically consistent boundary conditions, provides a quantitatively accurate description of propulsion in soft filamentary swimmers. Our findings highlight the importance of boundary conditions as an important ingredient in elasto-hydrodynamic locomotion and provide a framework for designing and optimizing synthetic micro-swimmers and soft robotic systems.

\begin{acknowledgments}   
Supported by U.S. National Science Foundation under Grant Number CBET-2401729 and the Brazilian National Council for Scientific and Technological Development (CNPq), Grant Number 409180/2023-8.
\end{acknowledgments}   


\begin{thebibliography}{23}%
\makeatletter
\providecommand \@ifxundefined [1]{%
 \@ifx{#1\undefined}
}%
\providecommand \@ifnum [1]{%
 \ifnum #1\expandafter \@firstoftwo
 \else \expandafter \@secondoftwo
 \fi
}%
\providecommand \@ifx [1]{%
 \ifx #1\expandafter \@firstoftwo
 \else \expandafter \@secondoftwo
 \fi
}%
\providecommand \natexlab [1]{#1}%
\providecommand \enquote  [1]{``#1''}%
\providecommand \bibnamefont  [1]{#1}%
\providecommand \bibfnamefont [1]{#1}%
\providecommand \citenamefont [1]{#1}%
\providecommand \href@noop [0]{\@secondoftwo}%
\providecommand \href [0]{\begingroup \@sanitize@url \@href}%
\providecommand \@href[1]{\@@startlink{#1}\@@href}%
\providecommand \@@href[1]{\endgroup#1\@@endlink}%
\providecommand \@sanitize@url [0]{\catcode `\\12\catcode `\$12\catcode `\&12\catcode `\#12\catcode `\^12\catcode `\_12\catcode `\%12\relax}%
\providecommand \@@startlink[1]{}%
\providecommand \@@endlink[0]{}%
\providecommand \url  [0]{\begingroup\@sanitize@url \@url }%
\providecommand \@url [1]{\endgroup\@href {#1}{\urlprefix }}%
\providecommand \urlprefix  [0]{URL }%
\providecommand \Eprint [0]{\href{}{} }%
\providecommand \doibase [0]{https://doi.org/}%
\providecommand \selectlanguage [0]{\@gobble}%
\providecommand \bibinfo  [0]{\@secondoftwo}%
\providecommand \bibfield  [0]{\@secondoftwo}%
\providecommand \translation [1]{[#1]}%
\providecommand \BibitemOpen [0]{}%
\providecommand \bibitemStop [0]{}%
\providecommand \bibitemNoStop [0]{.\EOS\space}%
\providecommand \EOS [0]{\spacefactor3000\relax}%
\providecommand \BibitemShut  [1]{\csname bibitem#1\endcsname}%
\let\auto@bib@innerbib\@empty

\bibitem [{\citenamefont {Gray}(1955)}]{gray1955movement}%
  \BibitemOpen
  \bibfield  {author} {\bibinfo {author} {\bibfnamefont {J.}~\bibnamefont {Gray}},\ }\bibfield  {title} {\bibinfo {title} {The movement of sea-urchin spermatozoa},\ }\href@noop {} {\bibfield  {journal} {\bibinfo  {journal} {J. Exp. Biol.}\ }\textbf {\bibinfo {volume} {32}},\ \bibinfo {pages} {775} (\bibinfo {year} {1955})}\BibitemShut {NoStop}%
\bibitem [{\citenamefont {Lighthill}(1975)}]{Lighthill1975}%
  \BibitemOpen
  \bibfield  {author} {\bibinfo {author} {\bibfnamefont {S.~J.}\ \bibnamefont {Lighthill}},\ }\href@noop {} {\emph {\bibinfo {title} {Mathematical biofluiddynamics}}}\ (\bibinfo  {publisher} {SIAM},\ \bibinfo {year} {1975})\BibitemShut {NoStop}%
\bibitem [{\citenamefont {Wiggins}\ \emph {et~al.}(1998)\citenamefont {Wiggins}, \citenamefont {Riveline},\ and\ \citenamefont {Goldstein}}]{wiggins1998trapping}%
  \BibitemOpen
  \bibfield  {author} {\bibinfo {author} {\bibfnamefont {C.~H.}\ \bibnamefont {Wiggins}}, \bibinfo {author} {\bibfnamefont {D.}~\bibnamefont {Riveline}},\ and\ \bibinfo {author} {\bibfnamefont {R.~E.}\ \bibnamefont {Goldstein}},\ }\bibfield  {title} {\bibinfo {title} {Trapping and wiggling: elastohydrodynamics of driven microfilaments},\ }\href@noop {} {\bibfield  {journal} {\bibinfo  {journal} {Biophys. J.}\ }\textbf {\bibinfo {volume} {74}},\ \bibinfo {pages} {1043} (\bibinfo {year} {1998})}\BibitemShut {NoStop}%
\bibitem [{\citenamefont {Wiggins}\ and\ \citenamefont {Goldstein}(1998)}]{wiggins1998flexive}%
  \BibitemOpen
  \bibfield  {author} {\bibinfo {author} {\bibfnamefont {C.~H.}\ \bibnamefont {Wiggins}}\ and\ \bibinfo {author} {\bibfnamefont {R.~E.}\ \bibnamefont {Goldstein}},\ }\bibfield  {title} {\bibinfo {title} {Flexive and propulsive dynamics of elastica at low {Reynolds} number},\ }\href@noop {} {\bibfield  {journal} {\bibinfo  {journal} {Phys. Rev. Lett.}\ }\textbf {\bibinfo {volume} {80}},\ \bibinfo {pages} {3879} (\bibinfo {year} {1998})}\BibitemShut {NoStop}%
\bibitem [{\citenamefont {Gray}\ and\ \citenamefont {Hancock}(1955)}]{gray1955propulsion}%
  \BibitemOpen
  \bibfield  {author} {\bibinfo {author} {\bibfnamefont {J.}~\bibnamefont {Gray}}\ and\ \bibinfo {author} {\bibfnamefont {G.~J.}\ \bibnamefont {Hancock}},\ }\bibfield  {title} {\bibinfo {title} {The propulsion of sea-urchin spermatozoa},\ }\href@noop {} {\bibfield  {journal} {\bibinfo  {journal} {J. Exp. Biol.}\ }\textbf {\bibinfo {volume} {32}},\ \bibinfo {pages} {802} (\bibinfo {year} {1955})}\BibitemShut {NoStop}%
\bibitem [{\citenamefont {Johnson}(1980)}]{Johnson1980}%
  \BibitemOpen
  \bibfield  {author} {\bibinfo {author} {\bibfnamefont {R.~E.}\ \bibnamefont {Johnson}},\ }\bibfield  {title} {\bibinfo {title} {An improved slender-body theory for {Stokes} flow},\ }\href@noop {} {\bibfield  {journal} {\bibinfo  {journal} {J. Fluid Mech.}\ }\textbf {\bibinfo {volume} {99}},\ \bibinfo {pages} {411} (\bibinfo {year} {1980})}\BibitemShut {NoStop}%
\bibitem [{\citenamefont {Lauga}\ and\ \citenamefont {Powers}(2009)}]{lauga2009hydrodynamics}%
  \BibitemOpen
  \bibfield  {author} {\bibinfo {author} {\bibfnamefont {E.}~\bibnamefont {Lauga}}\ and\ \bibinfo {author} {\bibfnamefont {T.~R.}\ \bibnamefont {Powers}},\ }\bibfield  {title} {\bibinfo {title} {The hydrodynamics of swimming microorganisms},\ }\href@noop {} {\bibfield  {journal} {\bibinfo  {journal} {Rep. Prog. Phys.}\ }\textbf {\bibinfo {volume} {72}},\ \bibinfo {pages} {096601} (\bibinfo {year} {2009})}\BibitemShut {NoStop}%
\bibitem [{\citenamefont {Lauga}(2020)}]{lauga2020fluid}%
  \BibitemOpen
  \bibfield  {author} {\bibinfo {author} {\bibfnamefont {E.}~\bibnamefont {Lauga}},\ }\href@noop {} {\emph {\bibinfo {title} {The fluid dynamics of cell motility}}},\ Vol.~\bibinfo {volume} {62}\ (\bibinfo  {publisher} {Cambridge University Press},\ \bibinfo {year} {2020})\BibitemShut {NoStop}%
\bibitem [{\citenamefont {Pak}\ \emph {et~al.}(2011)\citenamefont {Pak}, \citenamefont {Gao}, \citenamefont {Wang},\ and\ \citenamefont {Lauga}}]{pak2011high}%
  \BibitemOpen
  \bibfield  {author} {\bibinfo {author} {\bibfnamefont {O.~S.}\ \bibnamefont {Pak}}, \bibinfo {author} {\bibfnamefont {W.}~\bibnamefont {Gao}}, \bibinfo {author} {\bibfnamefont {J.}~\bibnamefont {Wang}},\ and\ \bibinfo {author} {\bibfnamefont {E.}~\bibnamefont {Lauga}},\ }\bibfield  {title} {\bibinfo {title} {High-speed propulsion of flexible nanowire motors: Theory and experiments},\ }\href@noop {} {\bibfield  {journal} {\bibinfo  {journal} {Soft Matter}\ }\textbf {\bibinfo {volume} {7}},\ \bibinfo {pages} {8169} (\bibinfo {year} {2011})}\BibitemShut {NoStop}%
\bibitem [{\citenamefont {Yu}\ \emph {et~al.}(2006)\citenamefont {Yu}, \citenamefont {Lauga},\ and\ \citenamefont {Hosoi}}]{yu2006experimental}%
  \BibitemOpen
  \bibfield  {author} {\bibinfo {author} {\bibfnamefont {T.~S.}\ \bibnamefont {Yu}}, \bibinfo {author} {\bibfnamefont {E.}~\bibnamefont {Lauga}},\ and\ \bibinfo {author} {\bibfnamefont {A.}~\bibnamefont {Hosoi}},\ }\bibfield  {title} {\bibinfo {title} {Experimental investigations of elastic tail propulsion at low {Reynolds} number},\ }\href@noop {} {\bibfield  {journal} {\bibinfo  {journal} {Phys. Fluids}\ }\textbf {\bibinfo {volume} {18}} (\bibinfo {year} {2006})}\BibitemShut {NoStop}%
\bibitem [{\citenamefont {Rodenborn}\ \emph {et~al.}(2013)\citenamefont {Rodenborn}, \citenamefont {Chen}, \citenamefont {Swinney}, \citenamefont {Liu},\ and\ \citenamefont {Zhang}}]{rodenborn2013propulsion}%
  \BibitemOpen
  \bibfield  {author} {\bibinfo {author} {\bibfnamefont {B.}~\bibnamefont {Rodenborn}}, \bibinfo {author} {\bibfnamefont {C.-H.}\ \bibnamefont {Chen}}, \bibinfo {author} {\bibfnamefont {H.~L.}\ \bibnamefont {Swinney}}, \bibinfo {author} {\bibfnamefont {B.}~\bibnamefont {Liu}},\ and\ \bibinfo {author} {\bibfnamefont {H.}~\bibnamefont {Zhang}},\ }\bibfield  {title} {\bibinfo {title} {Propulsion of microorganisms by a helical flagellum},\ }\href@noop {} {\bibfield  {journal} {\bibinfo  {journal} {Proc. Natl. Acad. Sci. U.S.A.}\ }\textbf {\bibinfo {volume} {110}},\ \bibinfo {pages} {E338} (\bibinfo {year} {2013})}\BibitemShut {NoStop}%
\bibitem [{\citenamefont {Liu}\ \emph {et~al.}(2025)\citenamefont {Liu}, \citenamefont {Chen},\ and\ \citenamefont {Zhang}}]{liu2025effective}%
  \BibitemOpen
  \bibfield  {author} {\bibinfo {author} {\bibfnamefont {B.}~\bibnamefont {Liu}}, \bibinfo {author} {\bibfnamefont {L.}~\bibnamefont {Chen}},\ and\ \bibinfo {author} {\bibfnamefont {J.}~\bibnamefont {Zhang}},\ }\bibfield  {title} {\bibinfo {title} {Effective and efficient modeling of the hydrodynamics for bacterial flagella},\ }\href@noop {} {\bibfield  {journal} {\bibinfo  {journal} {Phys. Fluids}\ }\textbf {\bibinfo {volume} {37}} (\bibinfo {year} {2025})}\BibitemShut {NoStop}%
\bibitem [{\citenamefont {Htet}\ and\ \citenamefont {Lauga}(2025)}]{htet2025load}%
  \BibitemOpen
  \bibfield  {author} {\bibinfo {author} {\bibfnamefont {P.~H.}\ \bibnamefont {Htet}}\ and\ \bibinfo {author} {\bibfnamefont {E.}~\bibnamefont {Lauga}},\ }\bibfield  {title} {\bibinfo {title} {Load-dependent resistive-force theory for helical filaments},\ }\href@noop {} {\bibfield  {journal} {\bibinfo  {journal} {Philos. Trans. R. Soc. A}\ }\textbf {\bibinfo {volume} {383}} (\bibinfo {year} {2025})}\BibitemShut {NoStop}%
\bibitem [{\citenamefont {Dasgupta}\ \emph {et~al.}(2013)\citenamefont {Dasgupta}, \citenamefont {Liu}, \citenamefont {Fu}, \citenamefont {Berhanu}, \citenamefont {Breuer}, \citenamefont {Powers},\ and\ \citenamefont {Kudrolli}}]{Dasgupta2013}%
  \BibitemOpen
  \bibfield  {author} {\bibinfo {author} {\bibfnamefont {M.}~\bibnamefont {Dasgupta}}, \bibinfo {author} {\bibfnamefont {B.}~\bibnamefont {Liu}}, \bibinfo {author} {\bibfnamefont {H.~C.}\ \bibnamefont {Fu}}, \bibinfo {author} {\bibfnamefont {M.}~\bibnamefont {Berhanu}}, \bibinfo {author} {\bibfnamefont {K.~S.}\ \bibnamefont {Breuer}}, \bibinfo {author} {\bibfnamefont {T.~R.}\ \bibnamefont {Powers}},\ and\ \bibinfo {author} {\bibfnamefont {A.}~\bibnamefont {Kudrolli}},\ }\bibfield  {title} {\bibinfo {title} {Speed of a swimming sheet in {Newtonian} and viscoelastic fluids},\ }\href {https://doi.org/10.1103/PhysRevE.87.013015} {\bibfield  {journal} {\bibinfo  {journal} {Phys. Rev. E}\ }\textbf {\bibinfo {volume} {87}},\ \bibinfo {pages} {013015} (\bibinfo {year} {2013})}\BibitemShut {NoStop}%
\bibitem [{\citenamefont {Johnson}\ and\ \citenamefont {Brokaw}(1979)}]{Johnson1979}%
  \BibitemOpen
  \bibfield  {author} {\bibinfo {author} {\bibfnamefont {R.~E.}\ \bibnamefont {Johnson}}\ and\ \bibinfo {author} {\bibfnamefont {C.~J.}\ \bibnamefont {Brokaw}},\ }\bibfield  {title} {\bibinfo {title} {Flagellar hydrodynamics. a comparison between resistive-force theory and slender-body theory},\ }\href {https://doi.org/10.1016/S0006-3495(79)85281-9} {\bibfield  {journal} {\bibinfo  {journal} {Biophys. J.}\ }\textbf {\bibinfo {volume} {25}},\ \bibinfo {pages} {113} (\bibinfo {year} {1979})}\BibitemShut {NoStop}%
\bibitem [{\citenamefont {Friedrich}\ \emph {et~al.}(2010)\citenamefont {Friedrich}, \citenamefont {Riedel-Kruse}, \citenamefont {Howard},\ and\ \citenamefont {J{\"u}licher}}]{Friedrich2010}%
  \BibitemOpen
  \bibfield  {author} {\bibinfo {author} {\bibfnamefont {B.~M.}\ \bibnamefont {Friedrich}}, \bibinfo {author} {\bibfnamefont {I.~H.}\ \bibnamefont {Riedel-Kruse}}, \bibinfo {author} {\bibfnamefont {J.}~\bibnamefont {Howard}},\ and\ \bibinfo {author} {\bibfnamefont {F.}~\bibnamefont {J{\"u}licher}},\ }\bibfield  {title} {\bibinfo {title} {High-precision tracking of sperm swimming trajectories with nanometer scale resolution},\ }\href@noop {} {\bibfield  {journal} {\bibinfo  {journal} {J. Exp. Biol.}\ }\textbf {\bibinfo {volume} {213}},\ \bibinfo {pages} {1226} (\bibinfo {year} {2010})}\BibitemShut {NoStop}%
\bibitem [{\citenamefont {Kantsler}\ and\ \citenamefont {Goldstein}(2012)}]{Kantsler2012}%
  \BibitemOpen
  \bibfield  {author} {\bibinfo {author} {\bibfnamefont {V.}~\bibnamefont {Kantsler}}\ and\ \bibinfo {author} {\bibfnamefont {R.~E.}\ \bibnamefont {Goldstein}},\ }\bibfield  {title} {\bibinfo {title} {Fluctuations, dynamics, and the solitary wave of a swimming microorganism},\ }\href@noop {} {\bibfield  {journal} {\bibinfo  {journal} {Phys. Rev. Lett.}\ }\textbf {\bibinfo {volume} {108}},\ \bibinfo {pages} {038103} (\bibinfo {year} {2012})}\BibitemShut {NoStop}%
\bibitem [{\citenamefont {Keaveny}\ \emph {et~al.}(2013)\citenamefont {Keaveny}, \citenamefont {Walker},\ and\ \citenamefont {Shelley}}]{Keaveny2013}%
  \BibitemOpen
  \bibfield  {author} {\bibinfo {author} {\bibfnamefont {E.~E.}\ \bibnamefont {Keaveny}}, \bibinfo {author} {\bibfnamefont {S.~W.}\ \bibnamefont {Walker}},\ and\ \bibinfo {author} {\bibfnamefont {M.~J.}\ \bibnamefont {Shelley}},\ }\bibfield  {title} {\bibinfo {title} {Optimization of chiral structures for microscale propulsion},\ }\href@noop {} {\bibfield  {journal} {\bibinfo  {journal} {Nano Lett.}\ }\textbf {\bibinfo {volume} {13}},\ \bibinfo {pages} {531} (\bibinfo {year} {2013})}\BibitemShut {NoStop}%
\bibitem [{\citenamefont {Ramananarivo}\ \emph {et~al.}(2013)\citenamefont {Ramananarivo}, \citenamefont {Godoy-Diana},\ and\ \citenamefont {Thiria}}]{ramananarivo2013passive}%
  \BibitemOpen
  \bibfield  {author} {\bibinfo {author} {\bibfnamefont {S.}~\bibnamefont {Ramananarivo}}, \bibinfo {author} {\bibfnamefont {R.}~\bibnamefont {Godoy-Diana}},\ and\ \bibinfo {author} {\bibfnamefont {B.}~\bibnamefont {Thiria}},\ }\bibfield  {title} {\bibinfo {title} {Passive elastic mechanism to mimic fish-muscle action in anguilliform swimming},\ }\href@noop {} {\bibfield  {journal} {\bibinfo  {journal} {J. R. Soc. Interface}\ }\textbf {\bibinfo {volume} {10}} (\bibinfo {year} {2013})}\BibitemShut {NoStop}%
\bibitem [{\citenamefont {Biswas}\ \emph {et~al.}(2023)\citenamefont {Biswas}, \citenamefont {Huynh}, \citenamefont {Desai}, \citenamefont {Moss},\ and\ \citenamefont {Kudrolli}}]{biswas2023dynamics}%
  \BibitemOpen
  \bibfield  {author} {\bibinfo {author} {\bibfnamefont {A.}~\bibnamefont {Biswas}}, \bibinfo {author} {\bibfnamefont {T.}~\bibnamefont {Huynh}}, \bibinfo {author} {\bibfnamefont {B.}~\bibnamefont {Desai}}, \bibinfo {author} {\bibfnamefont {M.}~\bibnamefont {Moss}},\ and\ \bibinfo {author} {\bibfnamefont {A.}~\bibnamefont {Kudrolli}},\ }\bibfield  {title} {\bibinfo {title} {Dynamics of magnetoelastic robots in water-saturated granular beds},\ }\href@noop {} {\bibfield  {journal} {\bibinfo  {journal} {Phys. Rev. Fluids}\ }\textbf {\bibinfo {volume} {8}},\ \bibinfo {pages} {094304} (\bibinfo {year} {2023})}\BibitemShut {NoStop}%
\bibitem [{SI()}]{SI}%
  \BibitemOpen
  \href@noop {} {}\bibinfo {note} {See Supplemental Material at \url{URL-will-be-inserted-by-publisher} for characterization data, drag coefficients, analytical solution to EH model with Minimal BC, EH model shape comparisons, RFT propulsion calculations, and the list of videos.}\BibitemShut {Stop}%
\bibitem [{\citenamefont {Yang}\ \emph {et~al.}(2017)\citenamefont {Yang}, \citenamefont {Lu}, \citenamefont {Zhao},\ and\ \citenamefont {Kawamura}}]{Yang2017}%
  \BibitemOpen
  \bibfield  {author} {\bibinfo {author} {\bibfnamefont {K.}~\bibnamefont {Yang}}, \bibinfo {author} {\bibfnamefont {C.}~\bibnamefont {Lu}}, \bibinfo {author} {\bibfnamefont {X.}~\bibnamefont {Zhao}},\ and\ \bibinfo {author} {\bibfnamefont {R.}~\bibnamefont {Kawamura}},\ }\bibfield  {title} {\bibinfo {title} {From bead to rod: Comparison of theories by measuring translational drag coefficients of micron-sized magnetic bead-chains in {Stokes} flow},\ }\href {https://doi.org/10.1371/journal.pone.0188015} {\bibfield  {journal} {\bibinfo  {journal} {PLoS One}\ }\textbf {\bibinfo {volume} {12}},\ \bibinfo {pages} {1} (\bibinfo {year} {2017})}\BibitemShut {NoStop}%
\bibitem [{\citenamefont {Lauga}(2007)}]{Lauga2007}%
  \BibitemOpen
  \bibfield  {author} {\bibinfo {author} {\bibfnamefont {E.}~\bibnamefont {Lauga}},\ }\bibfield  {title} {\bibinfo {title} {Floppy swimming: Viscous locomotion of actuated elastica},\ }\href {https://doi.org/10.1103/PhysRevE.75.041916} {\bibfield  {journal} {\bibinfo  {journal} {Phys. Rev. E}\ }\textbf {\bibinfo {volume} {75}},\ \bibinfo {pages} {041916} (\bibinfo {year} {2007})}\BibitemShut {NoStop}%
\end{thebibliography}

%

\end{document}